\documentclass[conference]{IEEEtran}
\IEEEoverridecommandlockouts

\usepackage{cite}
\usepackage{amsmath,amssymb,amsfonts}
\usepackage{graphicx}
\usepackage[dvipsnames]{xcolor}
\def\BibTeX{{\rm B\kern-.05em{\sc i\kern-.025em b}\kern-.08em T\kern-.1667em\lower.7ex\hbox{E}\kern-.125emX}}  
\usepackage{array}
\usepackage{textcomp}
\usepackage{stfloats}
\usepackage{url}
\usepackage{verbatim}
\usepackage{enumitem}
\usepackage{stfloats}
\usepackage[bottom]{footmisc}
\usepackage{algpseudocode}
\usepackage[bottom]{footmisc}
\usepackage{mathtools}
\usepackage{subcaption}
\usepackage[linesnumbered,ruled,vlined]{algorithm2e}
\algrenewcommand{\algorithmiccomment}[1]{\hfill// #1}
\algnewcommand{\algorithmicor}{\textbf{ or }}
\algnewcommand{\OR}{\algorithmicor}
\usepackage{balance}

\begin{document}
\title{Optimization of Federated Learning's Client Selection for Non-IID Data Based on Grey Relational Analysis}
\author{
\IEEEauthorblockN{Shuaijun~Chen\textsuperscript{1},
    Omid~Tavallaie\textsuperscript{1},
    Michael~Henri Hambali\textsuperscript{1},\\
    Seid~Miad~Zandavi\textsuperscript{2},
    Hamed~Haddadi\textsuperscript{3},
    Nicholas~Lane\textsuperscript{4},
    Song~Guo\textsuperscript{5},
    Albert~Y.~Zomaya\textsuperscript{1}}
\IEEEauthorblockA{
    \hspace{1cm}\textsuperscript{1}School of Computer Science, The University of Sydney, Australia\\
    \textsuperscript{2}The Broad Institue of MIT and Harvard, USA\\
    \textsuperscript{3}Imperial College London, UK\\
    \textsuperscript{4}The University of Cambridge, UK\\
    \textsuperscript{5}The Hong Kong University of Science and Technology, HK\\
    \{sche5840, mham7549\}@uni.sydney.edu.au, \{omid.tavallaie, albert.zomaya\}@sydney.edu.au}
    szandavi@broadinstitute.org, h.haddadi@imperial.ac.uk, ndl32@cam.ac.uk, songguo@cse.ust.hk\\
}
\maketitle
\vspace{-7mm}
\begin{abstract}

Federated learning (FL) is a novel distributed learning framework designed for applications with privacy-sensitive data. Without sharing data, FL trains local models on individual devices and constructs the global model on the server by performing model aggregation. However, to reduce the communication cost, the participants in each training round are randomly selected, which significantly decreases the training efficiency under data and device heterogeneity. To address this issue, in this paper, we introduce a novel approach that considers the data distribution and computational resources of devices to select the clients for each training round. Our proposed method performs client selection based on the Grey Relational Analysis (GRA) theory by considering available computational resources for each client, the training loss, and weight divergence. To examine the usability of our proposed method, we implement our contribution on Amazon Web Services (AWS) by using the TensorFlow library of Python. We evaluate our algorithm's performance in different setups by varying the learning rate, network size, the number of selected clients, and the client selection round. The evaluation results show that our proposed algorithm enhances the performance significantly in terms of test accuracy and the average client's waiting time compared to state-of-the-art methods, federated averaging and Pow-d.
\end{abstract}


\section{Introduction}
Neural Networks (NN) have become a widely used approach in Computer Vision (CV) and Natural Language Processing (NLP) today. Traditionally, model training involved gathering task-related data and centralized training on high-performance data centers. However, recent advancements in mobile computing have made it possible to generate large amounts of data and run complex Machine Learning (ML) algorithms on mobile devices. For instance, tasks like autonomous driving and speech recognition often require training data sets reaching terabytes in size. Generating training data continuously on millions of mobile devices\cite{deng2020edge} makes centralized training approaches infeasible due to the significant increase in communication costs. Besides, centralized approaches cannot be applied to applications that use sensitive data (e.g., health applications), as they compromise user privacy by collecting and uploading user data into a centralized server. 

\begin{figure}[t]
  \centering
    \includegraphics[width=1\linewidth, height=47 mm]{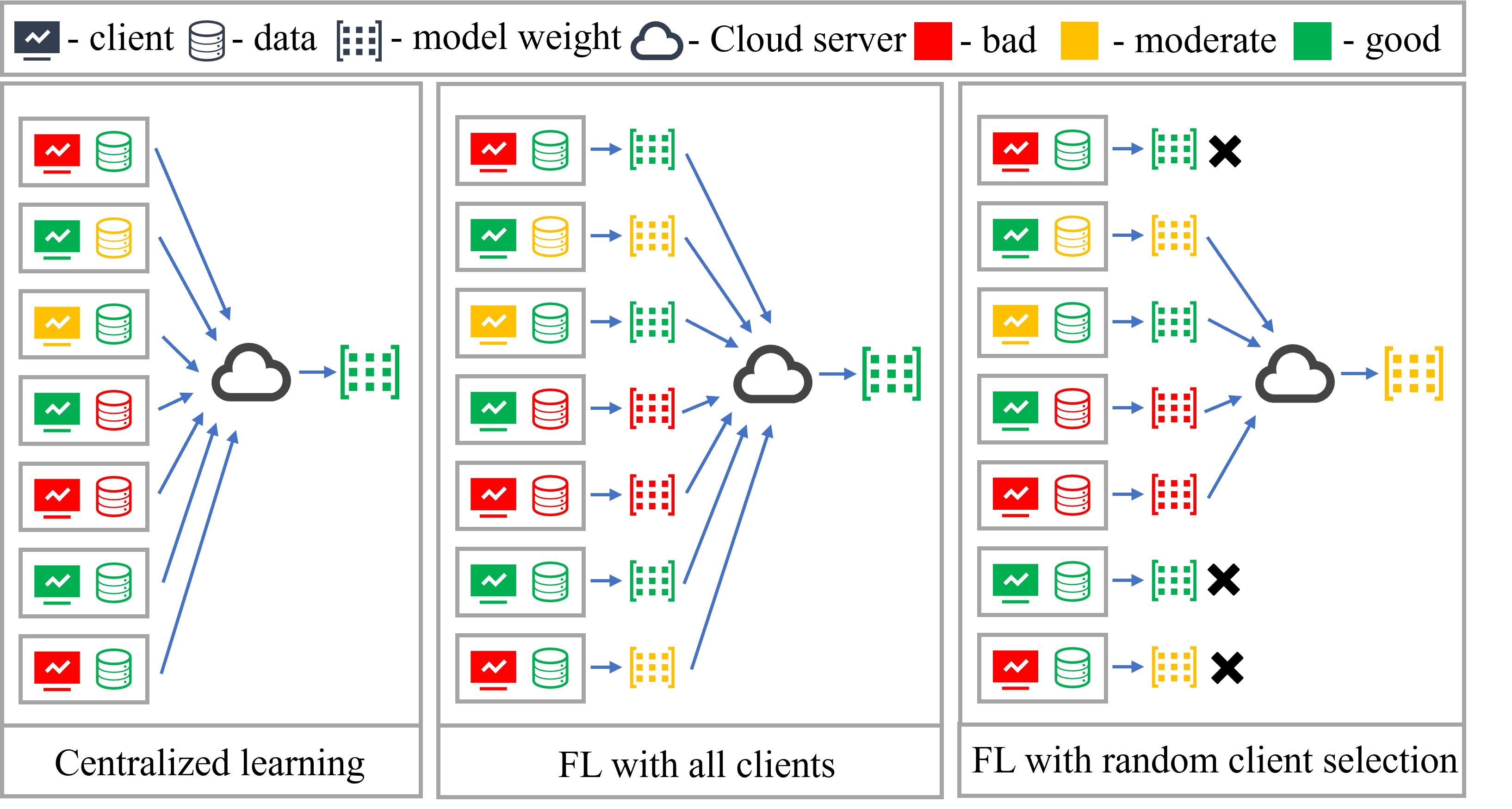}
  \caption{Participation of clients with heterogeneous data and hardware in the training process of 1) traditional centralized training, 2) federated learning, and 3) federated learning with client selection.}\vspace{-7 mm}
  \label{fig:problemStatement}
\end{figure}

\begin{figure}[t]
  \centering
    \includegraphics[width=75 mm, height=42 mm]{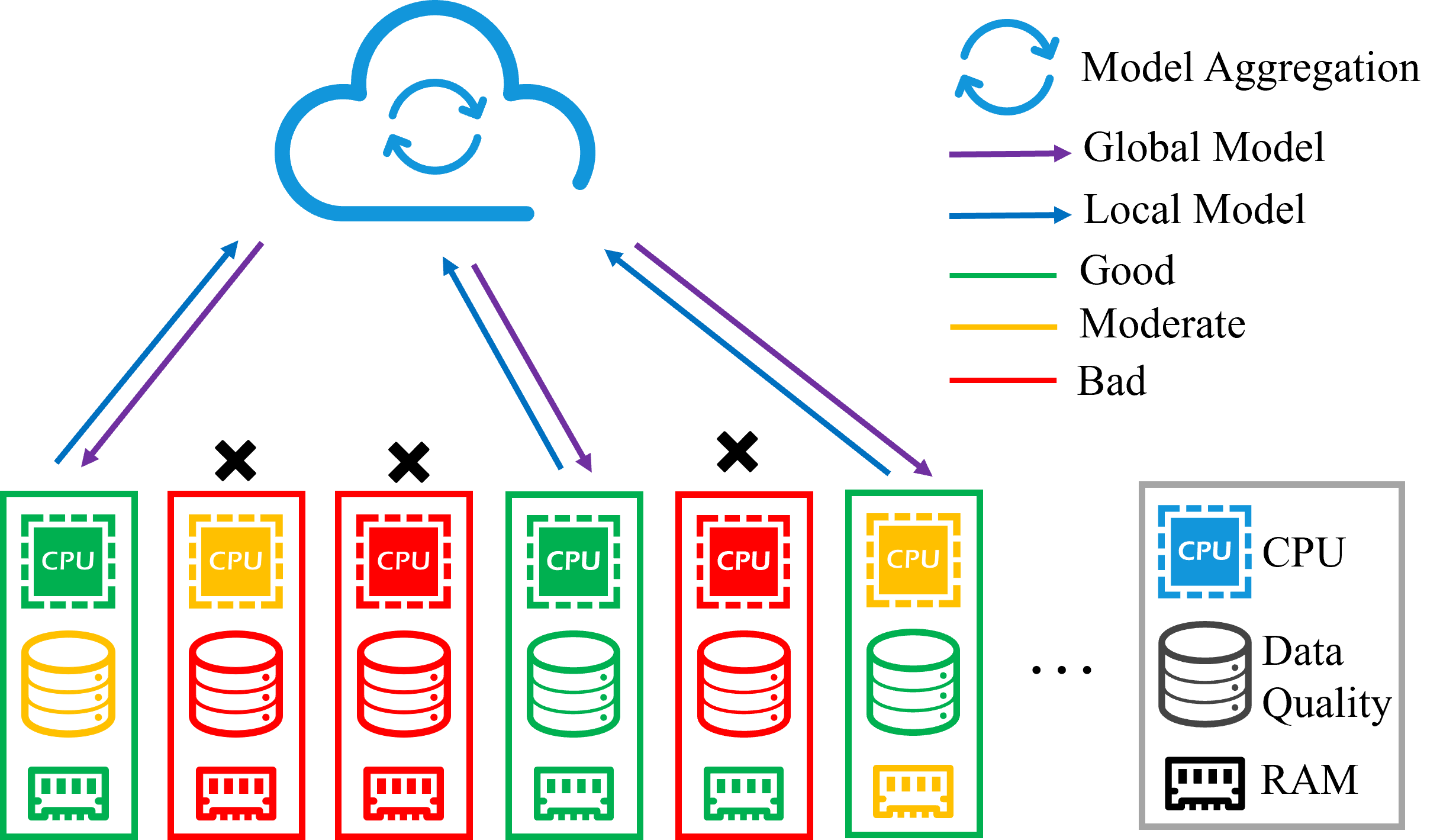}
  \caption{An optimal selection for 50\% of clients in a round of training by considering computational resources and data quality.}\vspace{-4 mm}
  \label{fig:optimalSelection}
\end{figure}

In 2017, McMahan et al. proposed Federated Learning (FL)\cite{mcmahan2017communication} as a privacy-aware distributed ML framework designed for decentralized NN training. Without sharing raw data, FL trains models locally on client devices and then uploads them to a central server for aggregation. Nevertheless, decentralized model training requires expensive communication costs on both the server and the client sides\cite{9220780}. For applications with a huge number of client devices (such as Google's Gboard\cite{xu-etal-2023-federated}), the vanilla Federated Averaging (FedAvg)\cite{mcmahan2017communication} randomly picks a limited number of {clients\cite{bonawitz2019towards}} in different training rounds to mitigate the training cost and communication overhead. Fig. \ref{fig:problemStatement} compares the participation of client devices in the training process of centralized and federating training (with and without client selection). As shown in this figure, under device and data heterogeneity, the random selection of client devices results in wasting computational resources and increasing the number of rounds caused by selecting resource-scare devices with poor data quality. In non-Identically and Independently Distributed (IID) data settings, the local dataset of a given client is not representative of the population distribution. As a result, the trained local models of two client devices may contribute considerably different in the aggregation process for building the global model{\cite{9464278}}. Under device heterogeneity, random client selection increases the average training time of an FL round. As an example, without considering computational resources, selecting client devices with less powerful processors increases the required time for a round of training as the server cannot start the aggregation before receiving local models from all clients. Furthermore, relying on random client selection may result in certain clients being excluded from the training process for extended periods of time. Therefore, labels can become excessively overfitted, discouraging clients and ultimately prompting them to lose interest in training. Fig. \ref{fig:optimalSelection} shows an example of the optimal solution for selecting 50\% of client devices when data quality and computational resources vary among different devices.

Based on the above-mentioned issues, designing an efficient client selection algorithm for FL is in great demand. To address this challenge, various approaches have been explored. For instance, reinforcement learning\cite{deng2021auction} models can be trained to help the server make efficient decisions, or game-theory approaches can be used to find an optimal solution in the Pareto frontier among participating clients. However, it's important to note that while these methods can be effective, they are challenging to be implemented, demanding significant computational resources, inflexible in certain scenarios, and reliant on specific evaluation metrics. 

In this paper, by considering data quality and computational resources of client devices, we employ the Grey Relational Analysis (GRA) theory to design FedGRA, a fair client selection algorithm for applications of FL with non-IID data. Our proposed method considers the client selection problem of FL as a grey system by defining customized metrics based on the available CPU and RAM resources of mobile devices, the training loss of local models, and the weight deviation between the local and global models. To guarantee fairness, FedGRA keeps the participation rate of each client device higher than a predefined threshold. The main contributions of our work are summarized as follows:
\begin{enumerate}[topsep=0 pt, partopsep=0 pt, wide=0 pt]
\item By conducting an extensive set of experiments in a real testbed, we find the most important criteria that impact the performance of FL's client selection algorithms under data and device heterogeneity.
\item We analyze the stability issue of FL client selection and show how data distribution and client participation rate affect the stability of the test accuracy. 
\item We introduce a lightweight and flexible method with low communication overhead to improve the accuracy results and reduce the average waiting time of an FL's round.
\item By using the TensorFlow library, we implement FedGRA in Python to compare its performance with state-of-the-art methods. To evaluate our contribution in a practical environment, we use 50 $t2$ Amazon Elastic Compute Cloud (EC2) instances with 4 different types of hardware specifications that reflect the computational resources of most mobile devices (16GB as the maximum RAM). 
\end{enumerate}

The remainder of this paper is organized as follows: section \ref{sectionrelated} discusses related work. Section \ref{sectionproblemstatement} explains the most important criteria that impact the performance of FL's client selection. Section \ref{sectionfedgra} introduces FedGRA and the metrics used for the client selection. Section \ref{sectiongra} shows how GRA theory can be used for finding the optimal solution. Section \ref{sectionevaluation} is about implementing FedGRA and state-of-the-art methods, evaluating their performance, and analyzing the results. Finally, section \ref{sectionconclusion} concludes this paper.

\section{Related Work}\label{sectionrelated}
McMahan et al. introduced FedAvg in \cite{mcmahan2017communication} as a decentralized and privacy-aware model training approach. The efficiency of FedAvg is notably impacted when only a subset of clients participate in each training round. The primary goals of client selection methods are to enhance model performance with limited clients and reduce communication costs. Common client selection approaches involve either 1) assessing and selecting clients based on specific metrics like hardware performance, communication costs, and training loss \cite{nishio2019client, cho2022client, liu2022distributed}, or 2) considering clients' contributions to the training process \cite{lim2020hierarchical,zeng2020fmore}. Selection strategies often encompass game theory \cite{li2022pyramidfl, lai2021oort, huang2020efficiency, hu2022federated}, as well as reinforcement learning \cite{deng2021auction}.

In FL scenarios with non-IID data, clients can have entirely different local data that does not represent the overall dataset distribution\cite{9835537}. As a result, the accuracy of the global model significantly diminishes compared to scenarios with IID data. Various methods have been developed to reduce the impact of data non-IIDness \cite{Li2020On} by considering the discrepancy between the optimal values of global and weighted local objective functions. To mitigate the influence of non-IID data, \cite{MLSYS2020_1f5fe839, 10.5555/3495724.3497520, pmlr-v119-karimireddy20a} constrain local updates within proximity of the initial global model. \cite{10.5555/3294996.3295196} shares and personalizes auxiliary tasks, and \cite{10.5555/3495724.3496024, 10.5555/3454287.3454819} learn the global task based on client's data to increase the generalization ability of the global model. In client selection methods, training loss serves as a crucial metric. \cite{johnson2018training} demonstrated that higher loss leads to a higher gradient. Based on this, \cite{li2022pyramidfl, lai2021oort, zhao2022dynamic, cho2022client} select clients with higher local loss. \cite{lim2020hierarchical, zeng2020fmore} assess client contributions by taking into account label significance, computational resources, and the local model's accuracy. Additionally, \cite{saha2022data} utilizes the standard deviation of data to represent the level of data non-IIDness.

Device disparities cause prolonged waits between high-performance clients and those with limited computational capabilities. To improve the training efficiency, \cite{nishio2019client} uses the client training time cost to represent the client's computational resources. Inconsistent with dropping clients out of the given budget, \cite{li2020federated} adjusts the local workload of clients and limits the unsatisfied updates from incapable devices. To totally remove the waiting time, \cite{zhou2021tea} allows clients to join or exit federated learning at any time and \cite{yu2023async, t2020personalized, tan2022towards, you2022triple, xie2020asynchronous} update the model asynchronously.

To increase the generalization of the global model, \cite{rodio2023federated} changes the weight between clients dynamically to make a balance between training speed and global model quality. Nevertheless, client devices with extremely limited computational performance may still not be able to participate in training for a long period. To improve the fairness of the client's participation in training, \cite{huang2020efficiency} proposes RBCS-F to guarantee that the probability of each client participation is higher than a threshold. Besides, this work adjusts the aggregated model's weights of the low-performance clients to improve their contribution to the global model. \cite{mohri2019agnostic} focuses on the worst-performing client and \cite{lyu2020collaborative} guarantees fairness based on client contribution to avoid the global model being over-fitted to a group of clients.

\section{Problem Statement}\label{sectionproblemstatement}
To find the most important factors that affect the convergence of the global model and the utilization of computational resources, we conduct extensive sets of experiments in our testbed by using 50 $t2$ AWS EC2 instances with 4 different hardware configurations. In this section, we explain the main factors and problems that impact the performance of client selection algorithms under data and device heterogeneity.\vspace{-1 mm}

\subsection{Clients with Heterogeneous Computational Resources}
\begin{figure}[t]
\begin{minipage}[t]{0.44\linewidth}
  \centering
    \includegraphics[width=43.7 mm, height=42.5 mm]{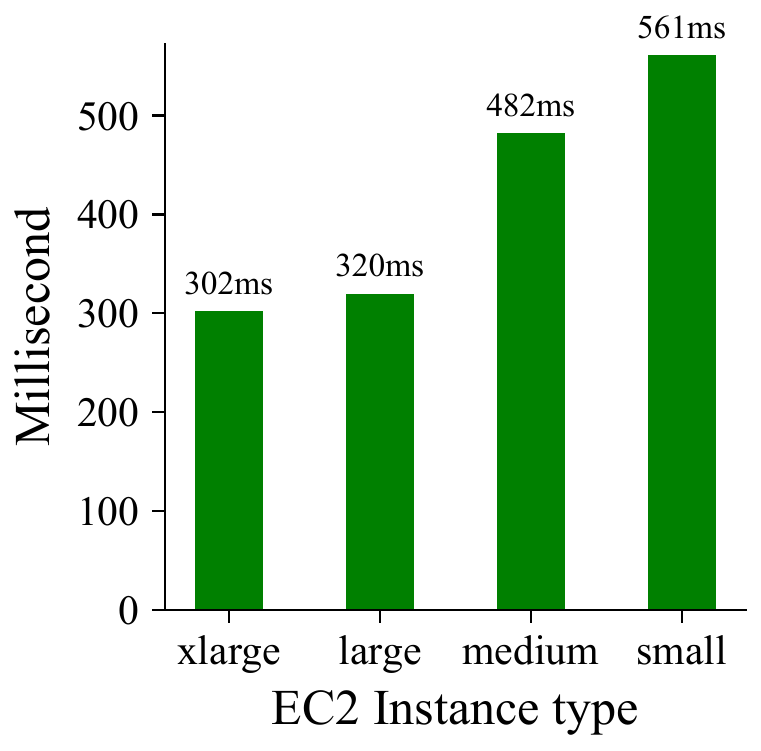}
  \caption{Training time for different Amazon EC2 instances.}
  \label{fig:ec2}
\end{minipage}
\hspace{0.15 cm}
\begin{minipage}[t]{0.44\linewidth} 
  \centering
    \includegraphics[width=43.7 mm, height=42.5 mm]{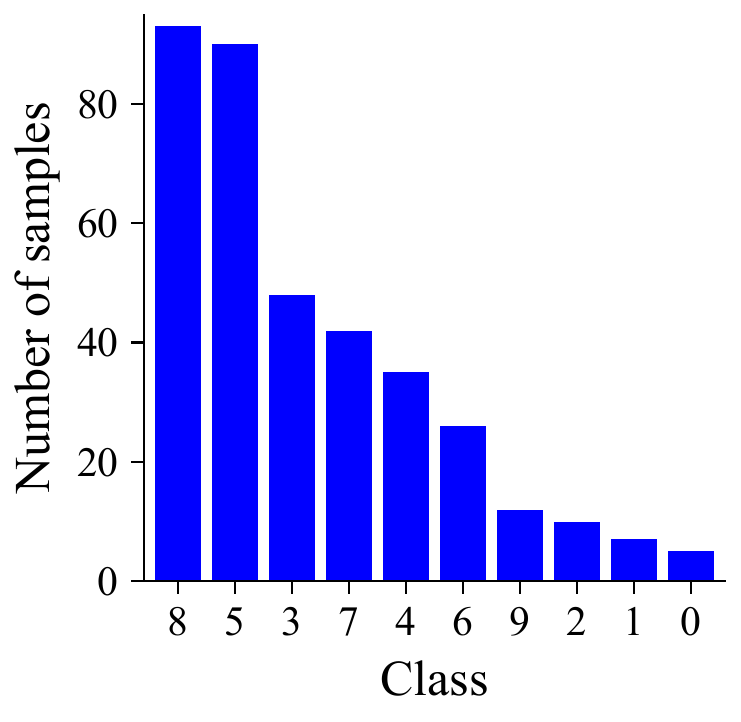}
  \caption{The long tail distribution for the number of samples per class.}
  \label{fig:longtail}
\end{minipage}    \vspace{-3 mm}    
\end{figure}

To reduce the energy consumption of client devices and communication cost, a small subset of clients with varying computational resources (including GPU, CPU, and RAM) is chosen for each training round. However, in vanilla federated learning, global model aggregation doesn't begin until updates are received from all client devices. As a result, the waiting time \textbf{(the time gap between finishing training the local model on clients with the most and the least computational resources and roughly the same amount of training data)} depends on devices with the least computational resources. Fig. \ref{fig:ec2} shows the average training time for four different AWS EC2 instance types selected based on the hardware configuration of most mobile devices. As shown in this figure, the training time for an \textit{xlarge} instance is around half of that for a \textit{small} instance. By selecting client devices with more powerful computational resources (such as \textit{xlarge} instances), more rounds of FL training could be run in a fixed period of time.

\begin{figure}[b]
    \centering
    \begin{subfigure}{0.22\textwidth}
       \centering
        \includegraphics[width=45 mm, height=42.5 mm]{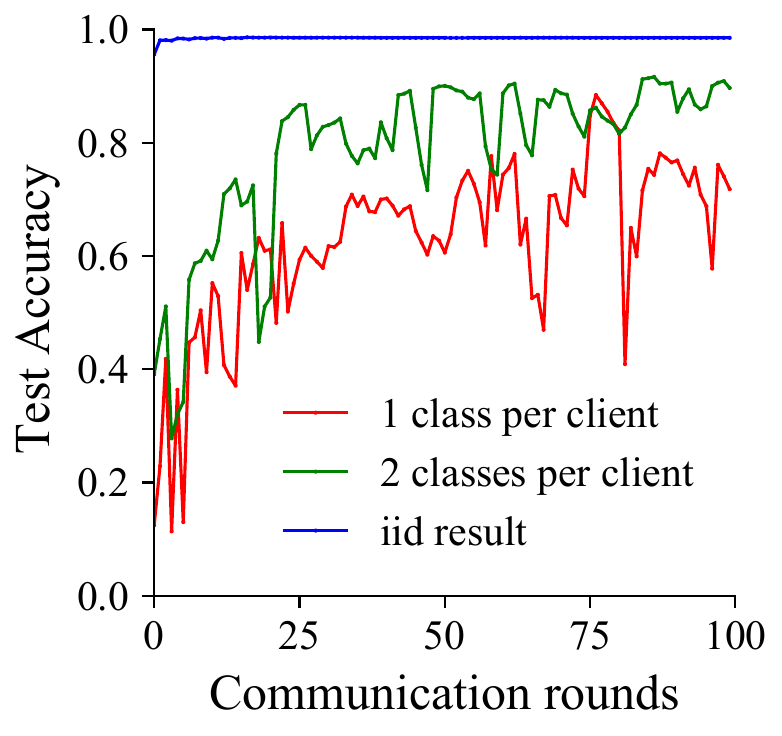}
        \caption{Varied data distribution.}
        \label{fig:mainmetrics-a}
    \end{subfigure}
    \hspace{1mm}
    \begin{subfigure}{0.22\textwidth}
    	\centering
        \includegraphics[width=45 mm, height=42.5 mm]{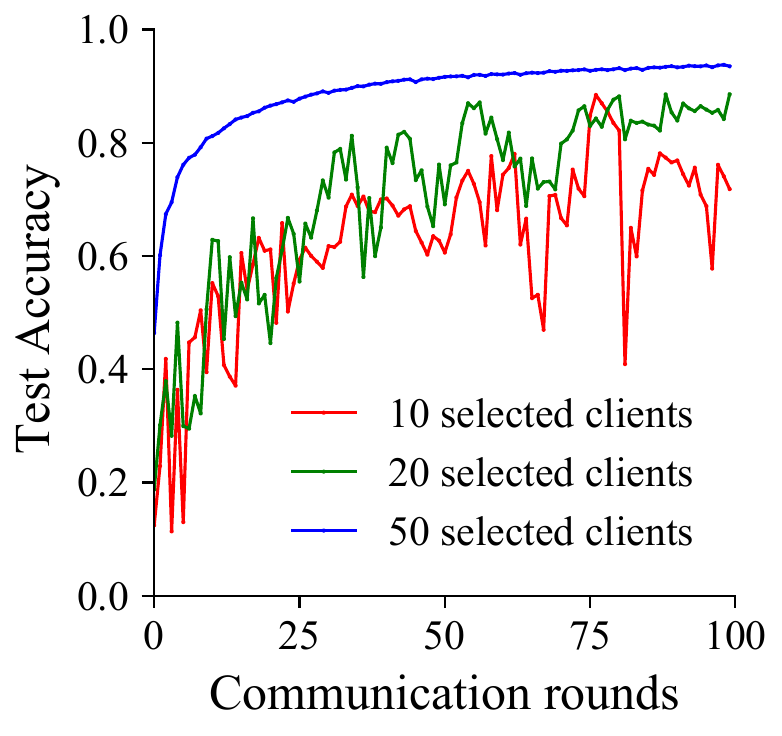}
        \caption{Varied number of clients.}
        \label{fig:mainmetrics-b}
    \end{subfigure}
    \caption{Evaluating FedAvg performance under varied data distribution and the number of selected clients.}
    \label{fig:mainmetrics}  \vspace{-7 mm}
\end{figure}
\subsection{Non-IID Data}
\begin{figure}[t]
  \centering
    \includegraphics[width=0.95\linewidth]
    {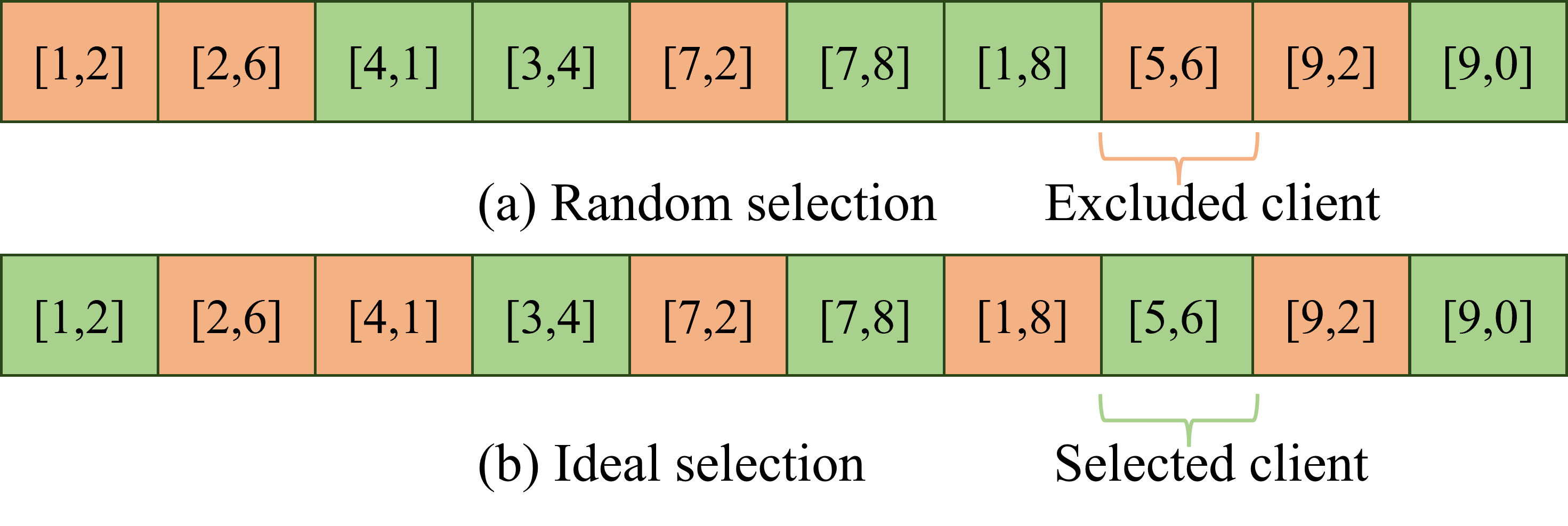} \vspace{-3 mm}
  \caption{Ideal client selection vs. random client selection. Boxes in green and orange indicate the selected clients and the clients excluded from training, respectively. Compared with the random selection, data distribution in ideal selected clients is more balanced, with a lower degree of non-IIDness.}
  \label{fig:idealselection}\vspace{-6 mm}
\end{figure}

In non-IID data scenarios of FL, clients have training data for different classes. As an example, Fig. \ref{fig:longtail} shows a non-IID data distribution for the image classification problem of MNIST dataset \cite{yang2020rethinking, zhang2023deep} under random client selection. As shown in this figure, the whole distributed data set used in a round of training is unbalanced (different classes have varied numbers of samples), following a long-tail distribution\cite{zhang2023deep, buda2018systematic}. In non-IID data scenarios, some classes being excessively over-fitted in one training round while the entire set of all classes are failed to be covered. For instance, in Fig. \ref{fig:longtail}, the number of samples for class 8 is many more than the sum of the numbers of samples for classes 0, 1, and 2. In this setting, by reducing the number of selected clients, more training rounds are required to achieve the same test accuracy. To assess this problem, we vary the data distribution and the number of selected clients in our experiments. As shown in Fig. \ref{fig:mainmetrics-a}, the test accuracy of the global model drops significantly as the number of classes assigned to each client decreases. When each client device has data for all classes (IID data), after around 2 rounds, the global model's accuracy reaches 98\%. However, for experiments with two different classes assigned to each client, the accuracy fluctuates widely in different rounds, and after 22 rounds it reaches to 80\%. These results indicate that distributing data with a higher level of non-IIDness among clients makes the convergence rate slower and increases accuracy fluctuation. Fig. \ref{fig:mainmetrics-b} shows the performance of FedAvg for different numbers of selected clients. When 10 clients are selected for training, it takes 75 rounds to reach around 80\% accuracy, which is 66 rounds more than that for the case all clients participate in training. Fig. \ref{fig:idealselection} shows the ideal client selection for non-IID data compared to the random selection. As this figure shows, for the ideal client selection, data from all classes are used in a round of training. Although, random selection does not select any client for class 2. In addition, the random selection also may not select certain clients for multiple training rounds. This becomes problematic in the presence of non-IID data, as a biased client selection can amplify the skewness of training data. Consequently, the global model might perform well only for a subset of classes. \vspace{-4 mm}

\section{FedGRA}\label{sectionfedgra}
To address the above-mentioned problems, we introduce FedGRA in this section, a fair client selection method designed based on the GRA theory to cope with data and device heterogeneity. 

\subsection{Measurements of Device heterogeneity}
In mobile federated learning scenarios, mobile client devices are not equipped with dedicated GPUs. Hence, client devices with more CPU and RAM resources have faster and more stable training processes. Based on this, we evaluate the available CPU and RAM resources of each client device to enhance training efficiency. 

\subsubsection{CPU Metric}
A consumer-grade CPU with superior computational performance typically owns a higher core count and clock frequency. In this context, we measure the CPU performance of client devices by defining $m_i$ as the CPU metric of client $i$ based on the number of CPU cores and the CPU clock frequency as:
\begin{equation}
    m_{i} = c_{i}*f_{i}*(1-l^{CPU}_{i}),
    \label{CPU_metric}
\end{equation}
where $c_i$, $f_i$, and $l_i^{CPU}$ represent the CPU core count, CPU clock frequency, and CPU load, respectively. 
\subsubsection{RAM Metric}
Similar to the CPU metric, client devices must have enough available RAM space to load the data and the neural network model. Thus, having more adequate available RAM space is more beneficial for local training. We define the RAM metric of client $i$ as $r_{i}$ in (\ref{ram_metric}) where $r^{total}_i$ and $l^{RAM}_i$ indicate the device's total memory and RAM usage, respectively: 
\begin{equation}
    \setlength\abovedisplayskip{2pt}
    \setlength\belowdisplayskip{3pt}
    r_{i}= r^{total}_{i}*(1-l^{RAM}_i).
    \label{ram_metric}
\end{equation}
As personal mobile devices typically run training programs in parallel with other tasks, utilization rates of computational resources depend on the other applications that a user runs on the mobile device. As an example, between two mobile devices with the same hardware configuration, training data on the device running a 3D game takes longer than a mobile device that hosts the web browser application. Although predicting the exact available computational resources for training data at a specific time is impractical, the Exponential Weighted Moving Average (EWMA) can be used to estimate the load of CPU and RAM resources based on their usage in the current and previous rounds. For instance, the weighted average RAM usage of device $i$ at round $t$ can be estimated by: 
\begin{equation}
    \overline{r_{i}}(t) = \theta(r_{i}(t))+(1-\theta)(\overline{r_{i}}(t-1)), 
    \label{EWMA}
\end{equation}
where $\theta$ represents the smoothing factor used to consider the historical data for the RAM load in previous training rounds. In our experiments, we set the value of $\theta$ to $0.9$.\vspace{-1mm}

\subsection{Measurements of Data Heterogeneity}
To ensure enough data for most classes used in the training process, FedGRA employs local training loss and weight divergence to represent the training value for clients with well-fitted and poor-fitted data, respectively.
\subsubsection{Training Loss}
FedGRA selects clients with data well-fitted to the global model to ensure that most clients benefit. Long-tail distribution is one common distribution in most real FL scenarios\cite{feldman2020neural}. As illustrated in Fig. \ref{fig:longtail}, in this distribution, the number of samples for tail classes (data with low frequency) is considerably lower than those for head classes (data with high frequency). In this distribution, head classes have higher training values as they are more generalized throughout the FL system. As a result, the global model fits the head classes better (compared to other classes) due to training with a higher amount of data. We use the loss metric $h_i$ (for client $i$) to represent how the global model fits the client data. By considering this metric, FedGRA selects devices with lower loss values to guarantee data of most clients have been well-fitted to the global model. The calculation of $h_i$ is represented by (\ref{loss_metric}), where $E$ denotes the set of epochs:
\begin{equation}
    h_{i} = \sqrt{{\textstyle \sum_{e \in E}Loss_{e}^2}}.
    \label{loss_metric}
\end{equation}

\subsubsection{Weight Divergence}
FedGRA employs the weight divergence $d_i$ (for client $i$) to guarantee the global model covers the clients with tail classes. This metric is designed by using the $L2$ norm for the difference between the weights of the local and the global models. A higher $d_i$ indicates client's data produces a larger gradient, which means there is a higher difference between the client data and the data well-fitted to the global model. This metric represents the training value of tail classes since most of them are not well-fitted. (\ref{weight_distance}) shows the calculation of $d_i$ where $w_g$ and $w_i$ represent the weights of the global and the local models, respectively:
\begin{equation}
    d_{i} = {||w_g-w_i||_{2}}.
    \label{weight_distance}
\end{equation}
Algorithm \ref{alg:client} shows the process of sending updates from the client $i$ to the server in FedGRA.

\begin{algorithm}[t]
      $TotalLoss=0$ \\
      \For{$\forall e \in E$}
      {
      	 	\For{$\forall b \in B$}
       		{
       			$w_{i}\longleftarrow w_{i} - \eta \nabla \ell(w;b)$
       		}
       		Compute $Loss_{e}$ \\
       		$TotalLoss=TotalLoss + Loss_{e}^{2}$
      }
 
      $h_{i} = \sqrt{{\textstyle \sum_{e \in E}Loss_{e}^2}}$\\
      $m_{i}= c_{i}*f_{i}*(1-l_{i}^{CPU})$ \\
      $r_{i} = r_{i}*(1-l_{i}^{RAM})$ \\
      Send $[h_{i},m_{i},r_{i},w_{i}]$ to the server
    \caption{FedGRA procedure executes on the client. $h_{i},m_{i},r_{i},w_{i}, E, B$ represent the training loss, CPU metric, RAM metric, the model weights, epoch, and batch, respectively.}\label{alg:client}
\end{algorithm}\setlength{\textfloatsep}{2 mm}

\subsection{Mapping}
FedGRA considers several metrics from different dimensions. It facilitates fair and meaningful comparisons among these metrics by normalizing source data into the range $[0,1]$. Due to GRA's constraints, all metrics used in FedGRA must either deviate from or approach the target (optimum) value. FedGRA picks clients with lower training loss, higher weight divergence, CPU and RAM performance. Based on this, we use positive and negative correlated metrics ($p_{i}$ and $n_{i}$) and then normalize them in (\ref{positive_mapping}) and (\ref{negative_mapping}) into the range $[0,1]$ to define $p_{i}^{'}$ and $ n_{i}^{'}$ as:

\begin{equation}
    p_{i}^{'} = \frac{p_{i}-\min_{i \in C}(p_{i})}{\max_{i \in C}(p_{i})+\min_{i \in C}(p_{i})},  \vspace{-3 mm}
    \label{positive_mapping}
\end{equation}

\begin{equation}
    n_{i}^{'} = \frac{\max_{i \in C}(n_{i}) - n_{i}}{\max_{i \in C} (n_{i}) + \min_{i \in C} n_{i}}.
    \label{negative_mapping}
\end{equation}
\section{Client Selection based on the GRA theory} \label{sectiongra}
In this section, we introduce our proposed client selection method FedGRA, including how we observe the client's hardware performance and data usability, the principle of GRA, the Entropy Weight Method (EWM), and the procedure to calculate the Grey Relational Grade (GRG) for each client. Table I shows the main notations we use in our paper.\vspace{-1 mm}

\begin{table}[b]
\renewcommand{\arraystretch}{1.3} %
\centering
\caption{Declaration of GRA notations}
\begin{tabular}{|l|l|}
\hline
\textbf{Notations} & \textbf{Definition} \\
\hline
$M$         & set of all metrics           \\
\hline
$C$         & set of all clients not being selected at the current round \\
\hline
$\Delta_{\max}$       & absolute maximum value           \\
\hline
$\Delta_{\min}$       & absolute minimum value           \\
\hline
$\Delta_{i}^{k}$       & distance difference of client $i$ on metric $k$          \\
\hline
$\bar{x}_{*}^{k}$         &  highest value of the normalized metric $k$           \\
\hline
${x}_{i}^{k}$         &  value of metric $k$ for client $i$            \\
\hline
$w^{k}$         & grey relational coefficient weight of metric $k$            \\
\hline
$p_{i}^{k}$     & the weight of the indicator value of the metric $k$ for client $i$ \\
\hline
$E^{k}$     & information entropy of metric $k$ \\
\hline
$g_i$       & FedGRA correlation factor \\
\hline
$\rho$       & distinguishing coefficient           \\
\hline
$n$         & total number of clients selected at each round \\
\hline
$t$          & client selection round \\
\hline
$j^r$ & number of clients are selected by GRA at round \textit{$r$}\\
\hline
$F_{i}^{r}$ & fairness metric for client $i$ at round $r$ \\
\hline
$f$ & fairness increment \\
\hline
$b$ & fairness bound \\
\hline
\end{tabular}
\label{gra_notation}
\end{table}
\subsection{Information Observation}
Performing $t$-round client selection means that clients involved in training are re-selected every $t$ rounds when the server collects models and information about data quantity and the device's computational resources from all clients. In this round, each client performs local training with the minimum number of epochs, collects FedGRA's metrics, and sends them to the server.

\subsubsection{Normalization}
GRA compares metrics in different dimensions. After scaling metrics into range $[0,1]$, we normalize metrics to a uniform dimension to facilitate a simple and complete trend analysis. (\ref{normalization}) indicates the normalization method we use in FedGRA where $x_{i}^{k}$ represents the value of metric $k$ on client $i$, and $C$ is the set of all clients:

\vspace{-1 mm}
\begin{equation}
    \bar{x}_{i}^{k} = \frac{x_{i}^{k}}{\frac{1}{|C|}\sum_{i \in C} x_{i}^{k}}. 
    \label{normalization}
\end{equation}

\subsubsection{Grey Relational Coefficient (GRC)}
The GRC effectively quantifies the relevance of various metrics. In FedGRA, the metric deviation $\Delta_{i}^{k}$ is assessed using the absolute difference in (\ref{delta_current}), which captures the discrepancy between the highest metric value and the value for client $i$ in metric $k$. It is calculated by the optimal value $\bar{x}_{*}^{k}$ and the client's metric value $\bar{x}_{i}^{k}$ as:
\begin{equation}
    \Delta_{i}^{k} =  \left | \bar{x}_{*}^{k} - \bar{x}_{i}^{k}  \right |.    
    \label{delta_current}
\end{equation}
Then, FedGRA computes the maximum absolute difference ($\Delta_{\text{max}}$) and the minimum absolute difference ($\Delta_{\text{min}}$) for the entire GRA matrix by:
\begin{equation}
    \Delta_{\max} = \max(\max(\Delta_{i}^{k})_{\forall k \in M})_{\forall i \in C}),
    \label{delta_max}
\end{equation}
\vspace{-3 mm}
\begin{equation}
    \Delta_{\min} = \min(\min(\Delta_{i}^{k})_{\forall k \in M})_{\forall i \in C}).
    \label{delta_min}
\end{equation}

After computing the absolute maximum value, the absolute minimum value, and the distance difference, GRA integrates these three values to calculate the GRC for each metric of client $i$ by:\vspace{1 mm}
\begin{equation}
    \xi_{i}^k =\frac{\Delta_{\min}+\rho \Delta_{\max}}{\Delta_{i}^{k}+\rho \Delta_{\max}}.
    \label{GRC} \vspace{1 mm}
\end{equation}
where $\Delta_{\text{max}}$ and $\Delta_{\text{min}}$ are used in (\ref{GRC}) to make GRC dimensionless and facilitate a meaningful comparison of correlation between different clients in different metrics. Factor $\rho$ is used to control the difference level to ensure that GRC remains within an adjustable range. A smaller $\rho$ leads to a higher distinction among GRC values that makes $\xi$ more comparable (when $\rho$ is set to 0, the numerators in all GRCs are the same). Consequently, the GRC value becomes inversely proportional to the distance from the current client $i$ to the ideal client for the same metric (a larger value shows the better performance). The value of $\rho$ can be customized based on task requirements. In our experiment, we set $\rho$ to 0.5.
\subsubsection{Weights of GRC}
FedGRA uses mutually exclusive metrics loss and weight divergence. We adjust their priorities in the client selection process by using the Entropy Weight Method (EWM) to set the weight of the GRCs. For instance, EWM calculates the information entropy of each metric and gives higher weight to metrics with low information entropy as they are more comparable. At the beginning of the training, most of the clients have similar weight divergence due to the lack of enough training data for all classes. This problem is alleviated after performing several rounds of model aggregation at the server. Hence, in the first few rounds, EWM assigns more weight to the loss metric to ensure that the head classes can be trained. After a certain number of training rounds, training loss for most of the clients converges to a certain value and the information entropy of the loss will be higher than the weight divergence. Then, EWM considers a higher priority for the weight divergence and asks clients with data not fitted well to the global model to participate in training. This dynamic adjustment encourages the participation of clients whose classes are less adequately generalized by the global model. (\ref{p_ij}) and (\ref{entropy}) represents the calculation procedure for information entropy of metric $k$. In (\ref{p_ij}), $p_{i}^{k}$ is the weight of the indicator value for the normalized metric $k$ of client $i$. 

\begin{equation}
    p_{i}^{k} = \frac{\bar{x}_{i}^{k}}{\textstyle \sum_{i \in C}\bar{x}_{i}^{k}}, \;\;\;\;\;\; {\textstyle \sum_{i \in C}p_{i}^{k}} = {1},
    \label{p_ij}
\end{equation}

\begin{equation}
    E^k = -\frac{1}{ln(|C|)} {\textstyle \sum_{i \in C}p_{i}^{k}*ln(p_{i}^{k})}.
    \label{entropy}
\end{equation}
 $E^{k}$ is the information entropy of normalized metric $k$. A higher value of $E^{k}$ indicates a higher difference degree in the data for metric $k$. Hence, a higher weight is assigned to metric $k$ by:
\begin{equation}
    w^k = \frac{1-E^{k}}{ {\textstyle \sum_{k \in M} (1 - E^k)}}. 
    \label{EWM}
\end{equation}
\subsubsection{Grey Relational Grade}
After calculating $\xi^{k}$ and $w^k$ for each metric of clients, the Grey Relational Grade (GRG) for each client $i$ is calculated  by:

\begin{equation}
    g_i = \sum_{k \in M} \frac{1}{w^k} \xi _{i}^{k}
    \label{GRG}
\end{equation}
This metric represents the relevance between the client $i$ and the ideal client. A client with a higher GRG has a higher chance of being selected for training.

\subsection{Fairness Guarantee}
FedGRA guarantees that all clients are selected at least one time in a given number of rounds by using the fairness metric defined in (\ref{fairness_constraint}). This metric shows the number of rounds past from the last participation of the client in training. $f$ is designed to control the increment of fairness for clients in each round. When the fairness metric $F_{i}$ is higher or equal to a threshold $b$, we select the client in the next training period, and this metric is set to 1 for each client at the beginning of the training:

\begin{equation}
    F_{i}^{r+1} = \begin{cases}
    F_{i}^{r} + f, &\;\;\;\; \text{$F_{i}^{r}+f<b,$} \\
    1, &\;\;\;\; \text{else.}
    \end{cases}
    \label{fairness_constraint}
\end{equation}

\begin{figure}[t]
  \centering
    \includegraphics[width=1\linewidth]{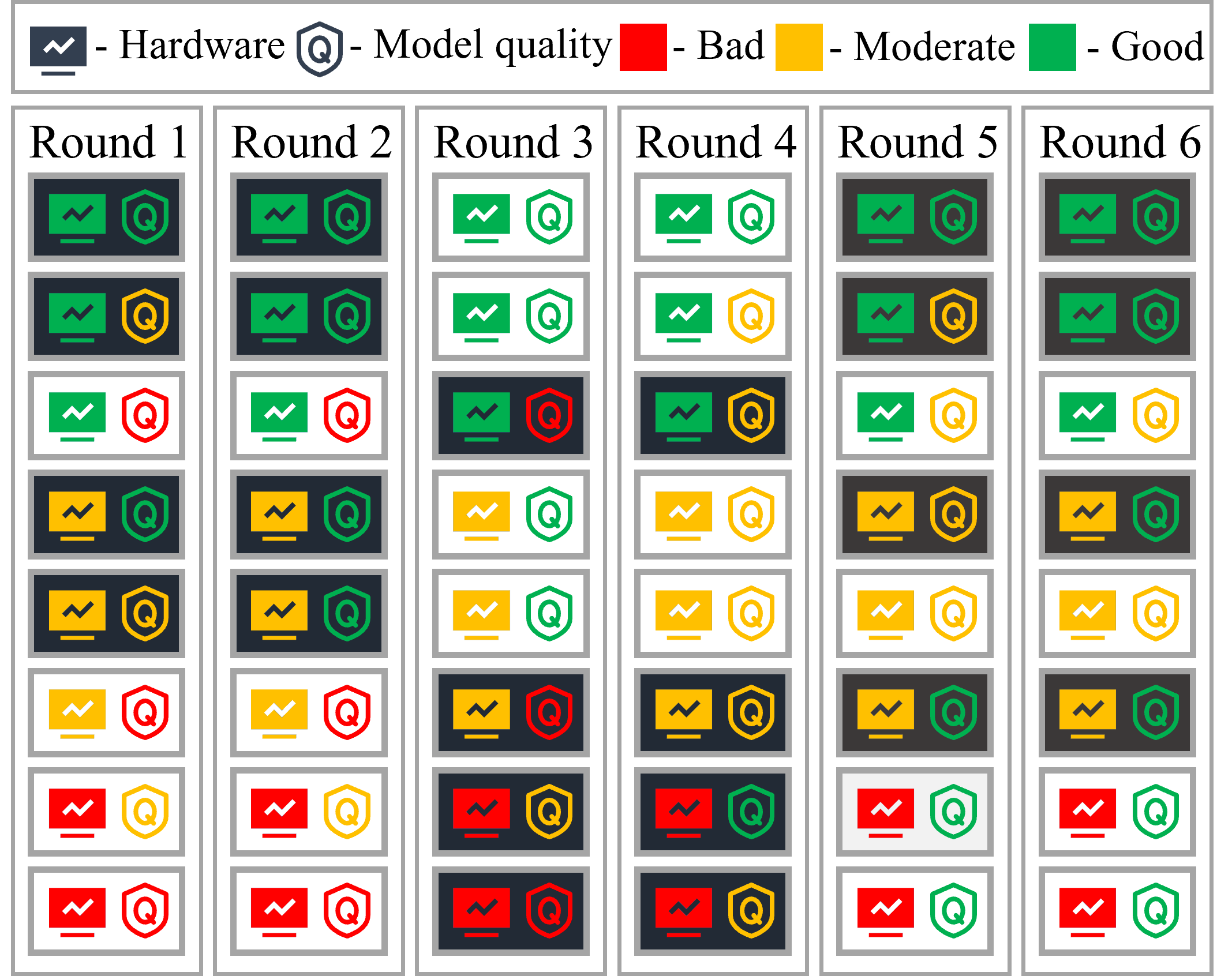}
  \caption{The result of running FedGRA for selecting 50\% of 8 clients.}
  \label{fig:selectionStrategy}
\end{figure}

Fig. \ref{fig:selectionStrategy} shows an example of running FedGRA for selecting 50\% of clients when the client selection round is 2 (i.e., the client selection is done once in two rounds of training).

\subsection{FedGRA Ranking}
After computing the FedGRA correlation factor $g_i$ for $\forall i \in C$, we sort the clients according to $g_i$ from highest to lowest values. The top $k\%$ of clients are selected to participate in the federated averaging procedure. Algorithm \ref{alg:server} shows the process of running FedGRA on the server for selecting $j$ clients.

\begin{algorithm}[t!]
\SetAlgoLined
\DontPrintSemicolon
\For{ each round $t$}  
 {
   \If{$t\mod t_{select}=0$}
   {  
       \text{Receive client information}\newline
       \textit{\color{Green}{\# Perform grey relational analysis}}\\
       \For{  $\forall k \in  M$ }
       {
      	    \text{Find} $\bar{x}_{*}^{k}$ \\
        	\For{$\forall i \in  C$}
       		 {
     		     $\Delta_{i}^{k} =  \left | \bar{x}_{*}^{k} - \bar{x}_{i}^{k}  \right |$    \\
        	 }
        }

       \For{  $\forall k \in  M$ }
       {
        	\For{$\forall i \in  C$}
       		 {
     				 $p_{i}^{k} = \frac{\bar{x}_{i}^{k}}{\sum_{i \in C}\bar{x}_{i}^{k}}$ \\       				
        	 }
        	  $E^{k} = -\frac{1}{ln(|C|)} { \sum_{i \in C}p_{i}^{k}*ln(p_{i}^{k})}$ \\
        }
       \For{ $\forall k \in M$}
       {
          $w^{k} = \frac{1-E^{k}}{ {\sum_{k \in M} (1 - E^k)}}$\\
       }
    
      \textit{\color{Green}{\# Absolute maximum and minimum values}} \\
      $\Delta_{\max} = \max(\max(\Delta_{i}^{k})_{\forall k \in M})_{\forall i \in C})$\\
      $\Delta_{\min} = \min(\min(\Delta_{i}^{k})_{\forall k \in M})_{\forall i \in C})$\\

     \For{ $\forall i \in C$}
     {
     	\textit{\color{Green}{\# Calculate grey relational grade}}
     	$ g_{i}= \sum_{\forall k \in M}\frac{1}{w^{k}} * \frac{\Delta_{\min}+\rho \Delta_{\max}}{\Delta_{i}(k)+\rho \Delta_{\max}} $\\
      }
        \textit{\color{Green}{\# Create the list of selected clients}}\\
        \If{$j^{r} \textgreater 0 $}
        {
        \text{Sort $g_{i}$ and select top $j^{r}$ clients} \\
        }
        \textit{\color{Green}{\# Update fairness}}\\
        \For{Each client $i$}
        	{	
        		\uIf{client $i$ is selected for training}
  				 {
  				 	$F_{i}^{r+1}= 1$
  				 }
  				\Else
  				{
  					$F_{i}^{r+1}= F_{i}^{r}+f$\\
  					\If{$F_{i}^{r+1} \geq b$}
  				 	{
  				 	\text{Select client $i$ for round $r+1$}\\
                         $F_{i}^{r+1}=1$ \\
                          $j^{r+1} = n - 1$
  				 	}  
  				}
        	}
        
   }
 }
    \caption{The FedGRA process of selecting $j^{r}$ clients at round $r$. $t_{select}$, $w$, $G$, and $E$ indicate the client selection round number, the local weight, the GRG metric, and the information entropy metric, respectively.}
\label{alg:server}
\end{algorithm}

\section{Implementation and Performance Evaluation}\label{sectionevaluation}
In our study, we evaluate the effectiveness of our proposed method using two different image classification datasets (MNIST and FMNIST) and compare its performance against the vanilla FedAvg\cite{mcmahan2017communication} and the advanced Pow-d\cite{cho2022client} methods. We use the TensorFlow library of Python for the implementation and run the code on 50 Amazon EC2 instances ($t2$ series) with 4 varied hardware configurations to test our contribution under heterogeneous devices. Our experiments included two neural network types (2NN and CNN) with the model structures shown in Table. \ref{table_2nn} and Table. \ref{table_CNN}, respectively. Additionally, we present varied training scenarios using the same dataset to demonstrate our method's stability and robustness, particularly in heterogeneous hardware settings and extreme non-IID situations. Table \ref{table:parameter_settings} shows the configuration parameters of our implementation.

\begin{table}[ht]
\centering
\caption{2NN model architecture.}
\begin{tabular}{|l|c|c|c|}
\hline
\textbf{Layer} & \textbf{Output Shape} & \textbf{Activation} & \textbf{Parameters} \\
\hline
Input & (784,) & None & 0 \\
\hline
Dense & (200,) & ReLU & 157,000 \\
\hline
Dense & (200,) & ReLU & 40,200 \\
\hline
Dense & (10,) & Softmax & 2,010 \\
\hline
\end{tabular}
\label{table_2nn}
\end{table}

\begin{table}[ht]
\caption{CNN model architecture.}
\centering
\begin{tabular}{|l|c|c|c|}
\hline
\textbf{Layer} & \textbf{Output Shape} & \textbf{Activation} & \textbf{Parameters} \\
\hline
Input & (28, 28, 1) & None & 0 \\
\hline
Conv2D & (28, 28, 32) & Sigmoid & 832 \\
\hline
MaxPooling2D & (14, 14, 32) & None & 0 \\
\hline
Conv2D & (14, 14, 64) & ReLU & 51,264 \\
\hline
MaxPooling2D & (7, 7, 64) & None & 0 \\
\hline
Flatten & (3136,) & None & 0 \\
\hline
Dense & (512,) & ReLU & 1,606,144 \\
\hline
Dense & (10,) & Softmax & 5,130 \\
\hline
\end{tabular}
\label{table_CNN}
\end{table}

\begin{table}[!t]
\caption{Configuration parameters of the experiments.}
\centering
\begin{tabular}{|l|c|c|c|}
\hline
Model architecture             & 2NN, CNN   \\
\hline
The total number of client devices                   & 50 \\
\hline
Learning rate                   & 0.1\\
\hline
Training round                  & 200\\
\hline
Epoch                           & 5\\
\hline
Batch size                      & 48\\
\hline
Number of selected clients      & 10\\
\hline
Selection rounds                & 5\\
\hline
Number of classes per client    & 1\\
\hline
Minimum participation rounds    & 6\\
\hline
\end{tabular}
\label{table:parameter_settings}
\end{table}

\subsection{Experiment setup}
In our experiment, we deployed our method on AWS to use AWS instances instead of code simulation to reflect the real effect of CPU and RAM performance on a practical FL training progress. Compared with the code simulation for the hardware performance, AWS instances provide realistic hardware environments that accurately reflect actual CPU and memory performance while code simulations are typically performed at the software level and hard to capture the details and characteristics at the hardware level (The specific impacts of CPU scheduling, timing, and concurrency). Moreover, the instances we used in our experiments reflect real hardware performance, as an example we use $t2$.small instances to simulate outdated devices like iPhone 8 and $t2$.xlarge to simulate powerful devices like OnePlus Ace Pro or Samsung S20 Ultra. In addition, we use another state-of-the-art method, Pow-d \cite{cho2022client}, to compare its performance with our contribution. The key idea of Pow-d is based on the power of $d$ choices load balancing strategy\cite{mitzenmacher2001power}. In the Pow-d method, $d$ clients are randomly sampled from the FL system. This approach gives priority to clients with higher local losses, aiming to accelerate the convergence of the global model more efficiently compared to a completely stochastic client selection process. This strategy effectively minimizes the chances of repeatedly involving the same client in training sessions and effectively reduces the complexity of searching for clients with higher local loss in large-scale FL systems. In general, Pow-d reduces the time cost search for clients with the highest local loss to balance time cost and convergence in the client selection process.

\begin{figure}[hb]
    \centering
    \includegraphics[width=90 mm, height=42 mm]{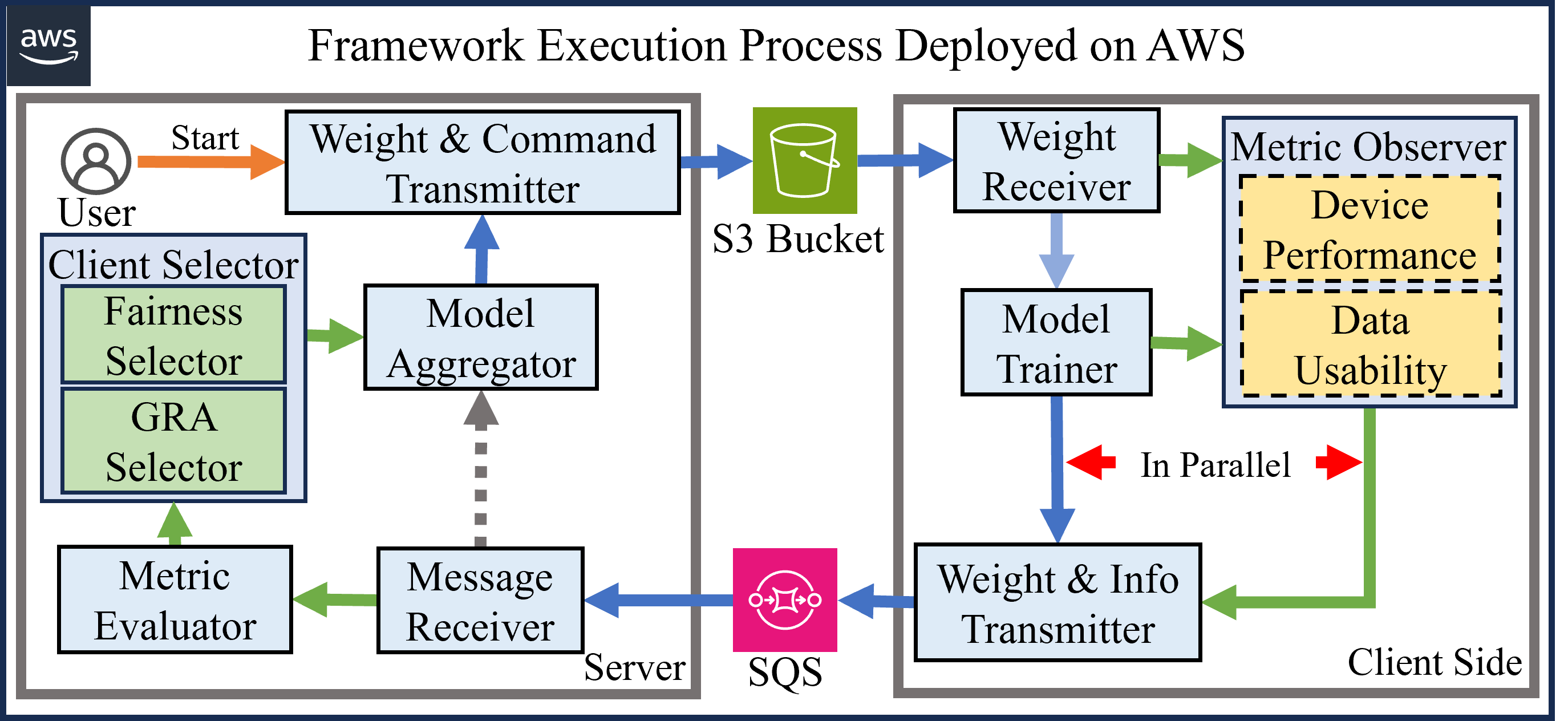}
  \caption{Implementation of FedGRA by using AWS. The blue, green, and grey lines show the procedures executed in every round, for the client selection, and between the client selection rounds, respectively.}
  \label{fig:enitre_procedure}
\end{figure}

\vspace{-1 mm}
\subsection{AWS Implementation}
In this section, we introduce how we set up our experiment on AWS. We utilize an AWS EC2 $t3$ instance as the server to aggregate the global model and send commands to clients. We manage storage and transmission between clients and the server by using AWS Simple Storage Service (S3) and Simple Queue Service (SQS) to ensure efficient communication and coordination in the FL process. Table. \ref{table:ec2_instance_specifications} and Table. \ref{table:table_t2_instance} show the AWS instance's configuration and the number of $t2$ instances (for clients) for CNN and 2NN models, respectively. To handle data distribution across client devices, Amazon S3 is used as the main storage for saving the data set and Python scripts required for running the model training on client devices. By using the AWS systems manager, we launch a bash command for all EC2 instances to download the files. To facilitate communication between the server and clients, we use Python sockets, Amazon S3, and Amazon SQS services that allow us to send, store, and receive messages from software components. These messages will contain the necessary information for either the server or the clients, such as locally trained weights, the globally aggregated model, and the number of data trained on each client. These messages are very large in terms of size. Therefore, we use Amazon S3 to store the messages and send the link of the message in the SQS message content. Utilizing Amazon SQS solves the problem of network congestion for a large number of clients. This is beneficial for the server to process multiple messages sequentially. The clients, on the other hand, do not need to utilize SQS as there is only one message from the server sent to every client. The server will only need to upload its message to S3 and then clients will receive it by checking the queue periodically. To avoid user dropout for instances with extremely limited resources (having 2GB of RAM), we select MNIST and FMNIST datasets for our experiments to ensure the stability and reproducibility of our results. User dropout is a critical factor for running FL on resource-limited devices that can affect results in different runs with the same configuration.

\begin{table}[t]
\caption{AWS $t2$ instance hardware specifications.}
\centering
\begin{tabular}{|l|c|c|c|c|}
\hline
\textbf{Metrics} & t2.small & t2.medium & t2.large & t2.xlarge \\
\hline
vCPUs & 1 & 2 & 2 & 4 \\
\hline
RAM(GB) & 2 & 4 & 8 & 16 \\
\hline
Physical Core & 1 & 2 & 2 & 4 \\
\hline
Clock speed & \multicolumn{4}{c|}{2.3 or {2.4} Ghz} \\
\hline
Turbo speed & \multicolumn{4}{c|}{3.0 Ghz}\\
\hline
OS             & \multicolumn{4}{c|}{Ubuntu} \\
\hline
\end{tabular}
\label{table:ec2_instance_specifications}
\end{table}

\begin{table}[t]
\caption{The number of instances used for each $t2$ instance type in experiments with different model architectures.}
\centering
\begin{tabular}{|l|c|c|c|c|}
\hline
\textbf{Model}   & t2.small & t2.medium & t2.large & t2.xlarge  \\
\hline
2NN               & 20       & 15       & 10     &5  \\
\hline
CNN               & 10       & 20       & 15     &5 \\
\hline
\end{tabular}
\label{table:table_t2_instance}
\end{table}

\subsection{AWS pipeline}
Fig. \ref{fig:enitre_procedure} represents the detailed task pipeline executed on clients and the server. The server's task includes assigning the client's data to make a non-IID scenario, aggregating model weight, distributing initial weights, and selecting clients based on different algorithms. The client's task performs local training, collecting the device's hardware and data metrics, and sending weight to the server. In the framework we deployed on AWS, clients and the server maintain a constant and dynamic interaction by using the AWS S3 bucket. They monitor a specified S3 bucket and by receiving a new JSON file, both the server and clients download and unpack it immediately. The server's bucket holds the instructional files for each client, which contain the tasks to be performed. Concurrently, the client's bucket contains clients' upload files containing their model weights and hardware metrics. Clients execute tasks according to the server's order, while the server aggregates clients' updates to refine a new global model. Subsequently, the server sends the command file to selected clients to participate in training in the next round.

To examine the performance of FedGRA under data heterogeneity, we use data with extreme non-IIDness to test our method (assigning data for only one class to each client device). To distribute the data set among client devices, first, we split the MNIST dataset into 10 different subsets while each subset has all data for one class. As an example, in MNIST, the subset for class 0 has 5923 samples. Then, we divide each data subset into slices with a predefined fixed size and assign one data slice to each client. As a result, for all slices belonging to the same class, only one slice has a slightly different size compared to the rest. \textbf{Each image in the origin MNIST dataset is assigned to only one client and we use the same data distributions for all experiments}.

\subsection{Evaluation Metrics}
$\bullet$ \textbf{Test Accuracy:} The test accuracy is defined as the accuracy on the test set for the aggregated global model on the server. We use the entire MNIST and FMNIST default test set with 10000 samples in our experiments.

$\bullet$ \textbf{Average waiting time:} The average waiting time is calculated as the mean duration that the client with the most powerful computational resources needs to wait for the client device equipped with the minimum computational resources to finish training.

\subsection{Evaluation Results}
\subsubsection{2NN and CNN}
\begin{figure}[!t]
    \centering
    \begin{subfigure}[b]{0.24\textwidth}
        \includegraphics[width=45 mm, height=42.5 mm]{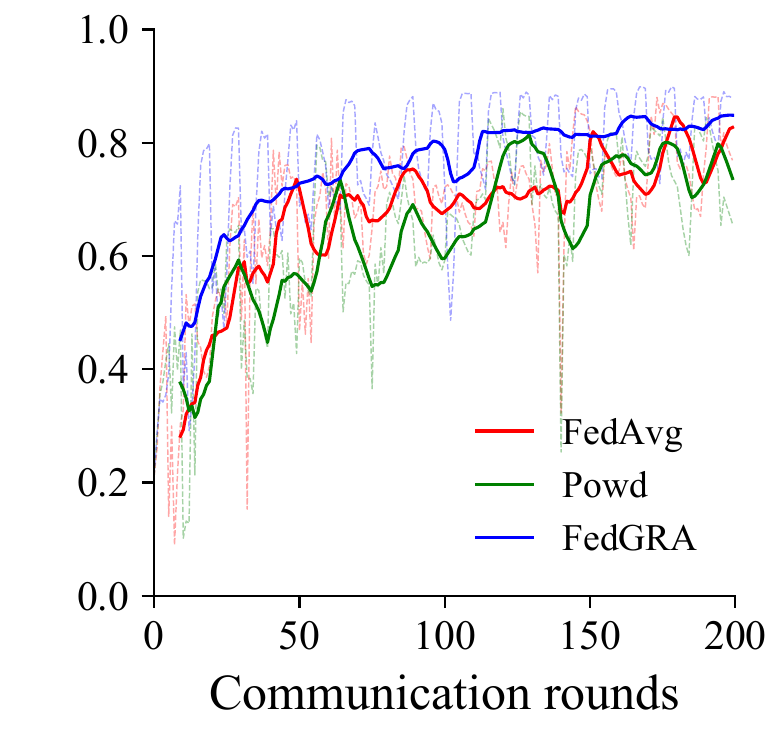}
        \caption{2NN results}
        \label{fig:mnist_2nn_baseline}
    \end{subfigure}
    \begin{subfigure}[b]{0.24\textwidth}
        \includegraphics[width=45 mm, height=42.5 mm]{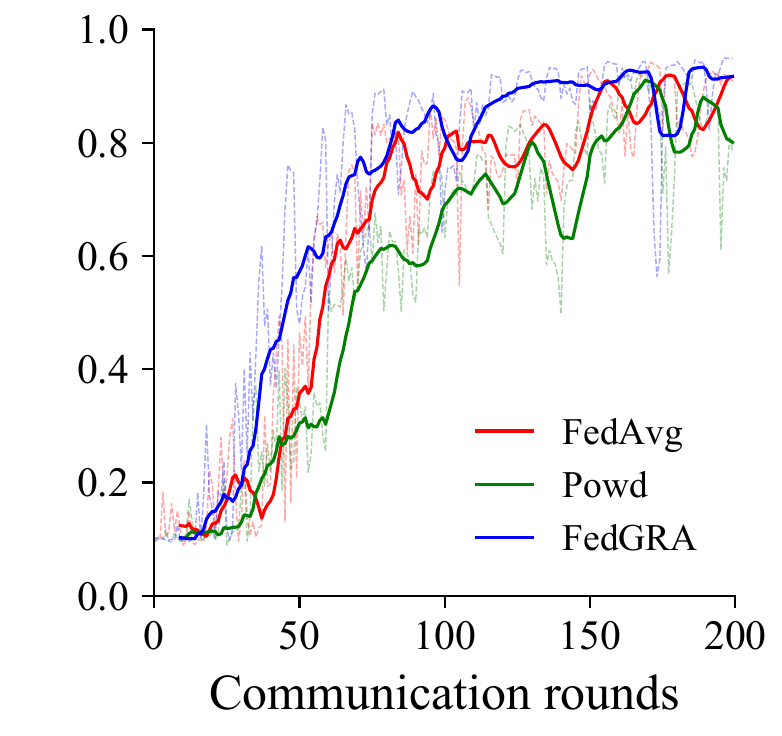}
        \caption{CNN results}
        \label{fig:mnist_cnn_baseline}
    \end{subfigure}
    \caption{Evaluation of FedGRA for 2NN and CNN models by using \textbf{MNIST} dataset.}
    \label{fig:mnist_baseline}
\end{figure}

\begin{figure}[!t]
    \centering
    \begin{subfigure}[b]{0.24\textwidth}
        \includegraphics[width=45 mm, height=42.5 mm]{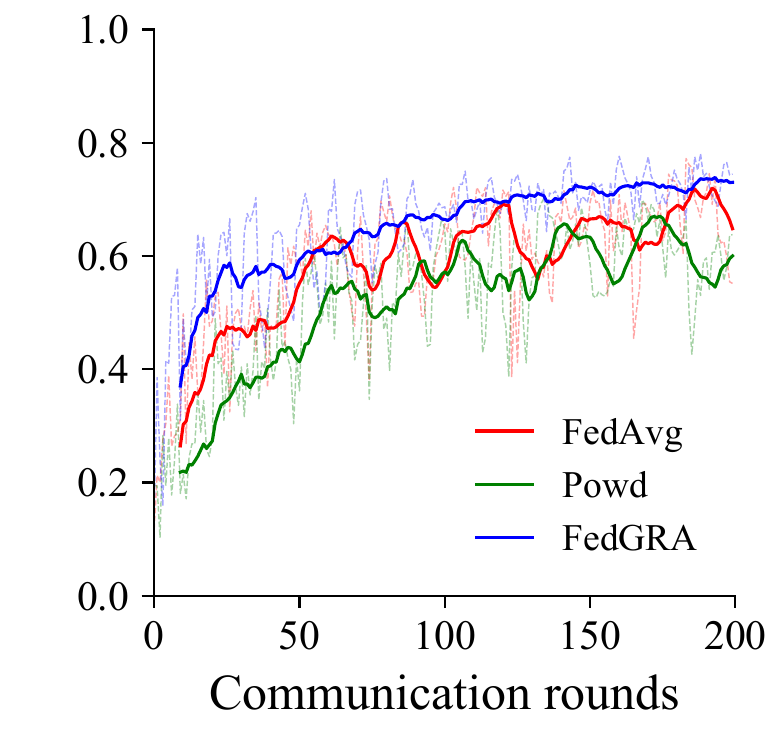}
        \caption{2NN results}
        \label{fig:fmnist_2nn_baseline}
    \end{subfigure}
    \begin{subfigure}[b]{0.24\textwidth}
        \includegraphics[width=45 mm, height=42.5 mm]{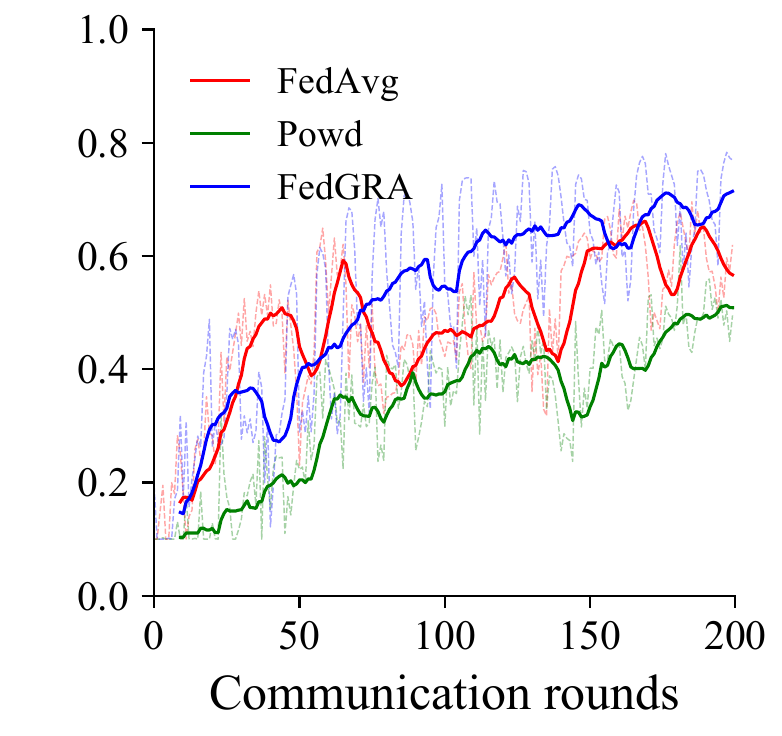}
        \caption{CNN results}
        \label{fig:famnist_cnn_baseline}
    \end{subfigure}
    \caption{Evaluation of FedGRA for 2NN and CNN models by using \textbf{FMNIST} dataset.}
    \label{fig:famnist_baseline}
\end{figure}

Fig. \ref{fig:mnist_baseline} and Fig. \ref{fig:famnist_baseline} demonstrate the test accuracy learning curves for MNIST and FMNIST datasets using 2NN and CNN models, respectively. In these experiments, an extreme case non-IID setting is applied, and to provide a clear comparison, we use a rolling average with a window size of 10 to smooth the data. The results indicate that FedGRA consistently performs better than both FedAvg and Pow-d in terms of stability and convergence.

In the MNIST 2NN experiments, FedGRA reaches 80\% accuracy with 19 training rounds, significantly faster than FedAvg and Pow-d, which require 62 and 57 rounds, respectively. This represents the \textbf{training rounds is reduced by 69\% compared with FedAvg.} In CNN experiments, to reach 90\% accuracy, FedGRA needs 79 rounds, 68 and 81 rounds less than FedAvg and Pow-d respectively. \textbf{It reduces the training rounds of FedAvg by 46\%.} The FMNIST 2NN and CNN experiments (Fig. \ref{fig:famnist_baseline}) further highlight FedGRA's efficiency. It achieves 70\% accuracy in 36 and 78 rounds of 2NN and CNN experiments, respectively, which shows a 65\% and 54\% reduction in the number of rounds required for FedAvg. For Pow-d, the reductions are even more substantial: 71\% for the 2NN model and 61\% for the CNN. Additionally, in both Fig. \ref{fig:mnist_2nn_baseline} and Fig. \ref{fig:mnist_cnn_baseline} FedGRA shows significantly lower test accuracy fluctuation than the other two methods with a more consistent and smoother rolled average accuracy.

\subsubsection{Average Waiting Time}
\begin{table}[b]
\caption{Average waiting time comparison}
\centering
\begin{tabular}{|l|c|c|c|}
\hline
 & FedAvg & FedGRA & Pow-d\\
\hline
CNN, MNIST Base & 6.24s & \textbf{4.09s} & 6.76s\\
\hline
CNN, FMNIST Base & 10.35s & \textbf{9.47s} & 10.87s\\
\hline
CNN, SR = 3 & 14.41s & \textbf{7.02s} & 9.36s\\
\hline
2NN, MNIST Base & 0.37s & 0.35s  & \textbf{0.27s} \\
\hline
2NN, FMNIST Base & 0.44s & \textbf{0.31}  & 0.45 \\
\hline
2NN, SR = 3 & \textbf{0.41s} & 0.49s  & 0.44s \\
\hline
2NN, LR = 0.125 & 0.29s & \textbf{0.24s} & 0.29s \\
\hline
2NN, 30 clients & 0.24s & 0.23s & 0.23s \\
\hline
\end{tabular}
\label{table:avg_wait_time}
\end{table}
This metric reflects the computational capacity gap between the most and least capable clients in training. A higher degree of hardware heterogeneity among the clients can significantly extend the waiting time for clients with high-performance resources. The average waiting time reflects how a client selection algorithm utilizes computational resources to mitigate the impact of device heterogeneity in the training process. This metric represents the degree of device heterogeneity involved in the training process and highlights the hardware capability gap among participant devices. As Table \ref{table:avg_wait_time} illustrates, for training MNIST by using CNN model, FedGRA's maximum waiting time is 48.7\% of that of FedAvg. However, for the experiment with 2NN, FedAvg demonstrates a slightly shorter waiting time than FedGRA owing to the relative simplicity of the 2NN model compared to CNN. These results indicate the problem of device heterogeneity becomes more critical by increasing the complexity of machine learning models.

\subsection{Robustness Analyse}
In this section, we present a robustness analysis for FedGRA under different hyper-parameter settings to evaluate the algorithm's robustness to the learning rate, total client number, and the number of selected clients under the data distribution with the high level of non-IIDness. Considering the heavy time cost to train the CNN model on extremely resource-limited instances, such as EC2 instances with only 2GB of RAM, we conduct experiments using the 2NN model described in Table.\ref{table_2nn} with MNIST and FMNIST dataset to guarantee the reliability and stability of our results.

\subsubsection{Scalability} 
\begin{figure}[ht]
    \centering
    \begin{subfigure}[b]{0.24\textwidth}
        \includegraphics[width=45 mm, height=42.5 mm]{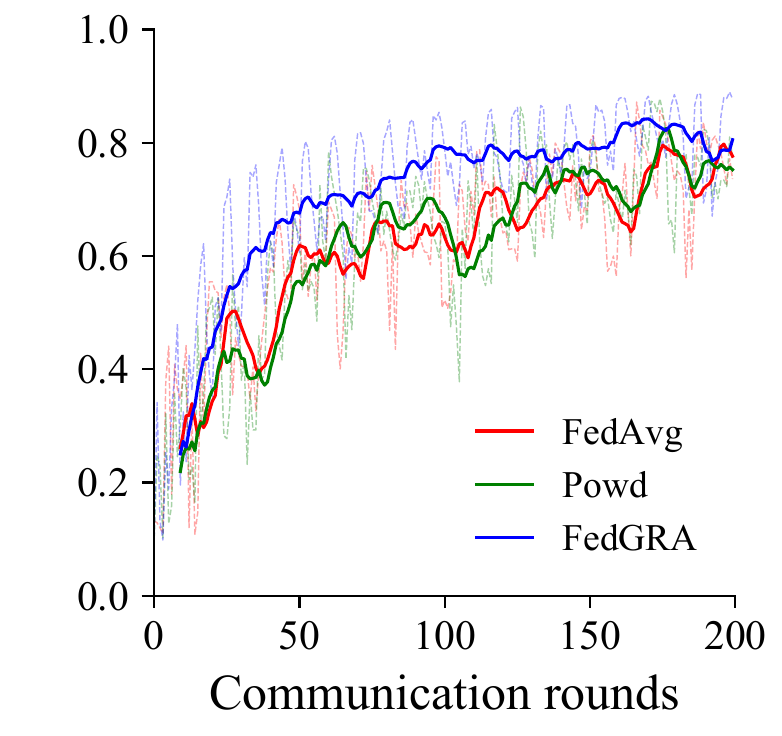}
        \caption{Total clients = 30}
        \label{fig:client_number_a}
    \end{subfigure}
    \begin{subfigure}[b]{0.24\textwidth}
        \includegraphics[width=45 mm, height=42.5 mm]{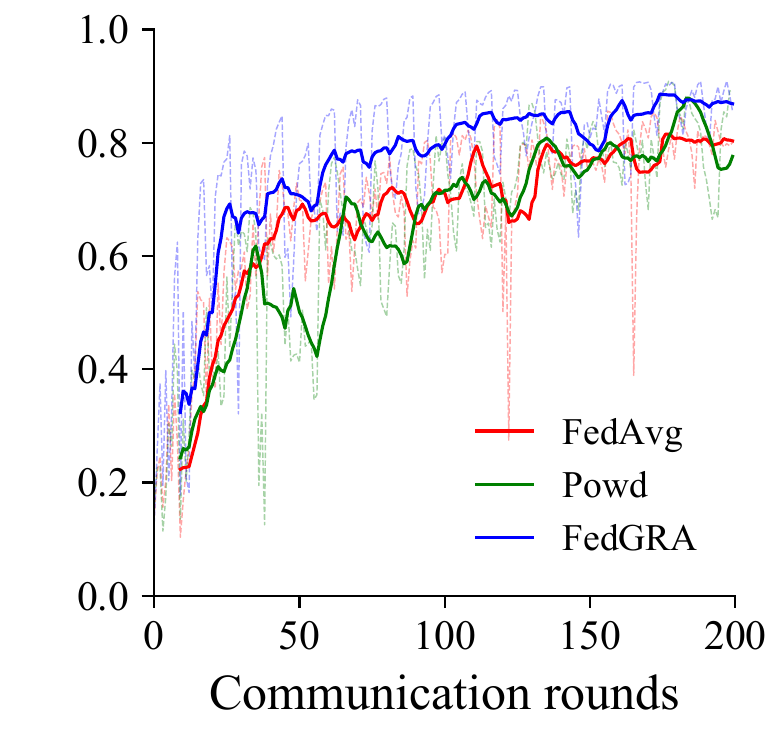}
        \caption{Total clients = 40}
        \label{fig:client_number_b}
    \end{subfigure}
    \caption{Evaluating the test set accuracy of FedGRA for different number of clients by using \textbf{MNIST} dataset.}
    \label{fig:client_number}
\end{figure}

\begin{figure}[ht]
    \centering
    \begin{subfigure}[b]{0.24\textwidth}
        \includegraphics[width=45 mm, height=42.5 mm]{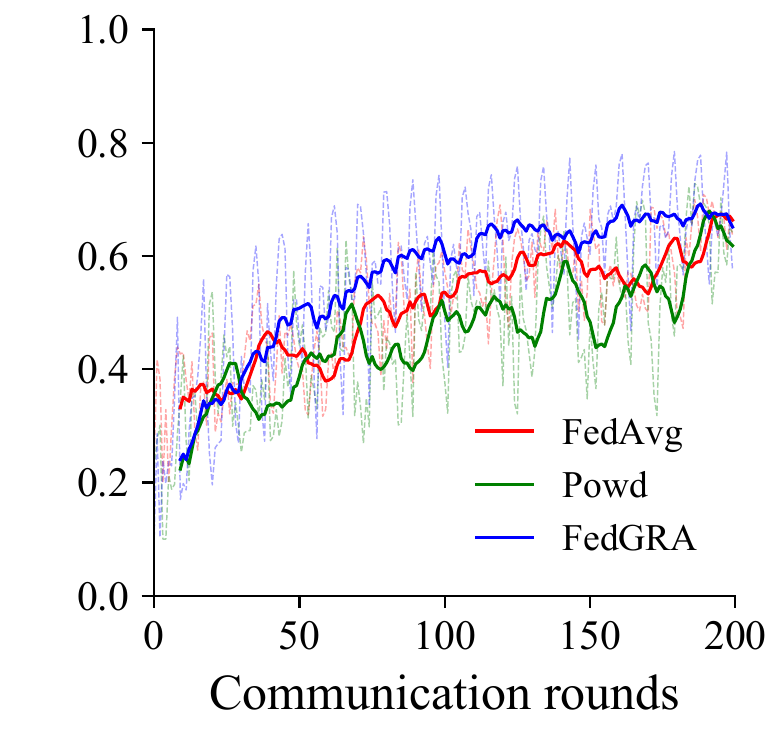}
        \caption{Total clients = 30}
        \label{fig:client_number_a_fmnist}
    \end{subfigure}
    \begin{subfigure}[b]{0.24\textwidth}
        \includegraphics[width=45 mm, height=42.5 mm]{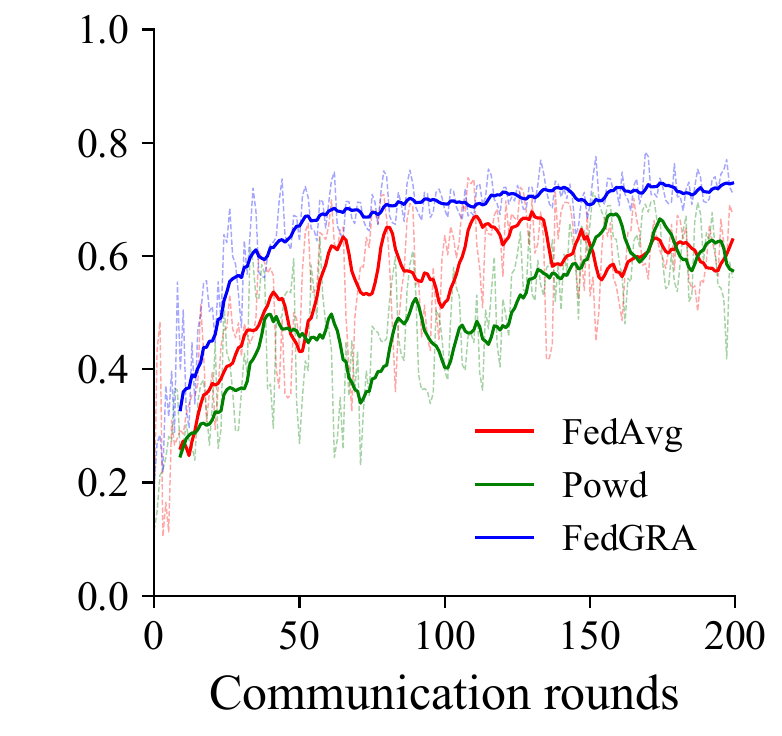}
        \caption{Total clients = 40}
        \label{fig:client_number_b_fmnist}
    \end{subfigure}
    \caption{Evaluating the test set accuracy of FedGRA for different number of clients by using \textbf{FMNIST} dataset.}
    \label{fig:client_number_fmnist}
\end{figure}

Fig. \ref{fig:client_number} and Fig. \ref{fig:client_number_fmnist} show the results of experiments with different total numbers of clients. In each experiment, 20\% of total clients are selected to participate in training for each round. As shown in Fig. \ref{fig:client_number} and Fig .\ref{fig:client_number_fmnist}, FedGRA converges faster and more stable compared to other methods. FedGRA reaches 80\% test accuracy for experiments with the sizes of 30 and 40 clients in 52 and 42 rounds for MNIST dataset, respectively (86 and 51 rounds less than that of FedAvg, and 74 and 58 rounds less than that of Pow-d). For FMNIST results, FedGRA reaches the $70\%$ accuracy in rounds 79 for 30 clients which is 110 and 105 rounds less than that of FedAVG and Pow-d. For the experiment with 40 clients, the accuracy of FedGRA reaches 70\% in 75 and 33 rounds less than the number of rounds required for training FedAvg and 76 for Pow-d. The rolled average accuracy of FedGRA is also higher than that for the other two methods in the vast majority of training rounds.

\subsubsection{Learning rate} 
\begin{figure}[t]
    \centering
    \begin{subfigure}[b]{0.24\textwidth}
        \includegraphics[width=45 mm, height=42.5 mm]{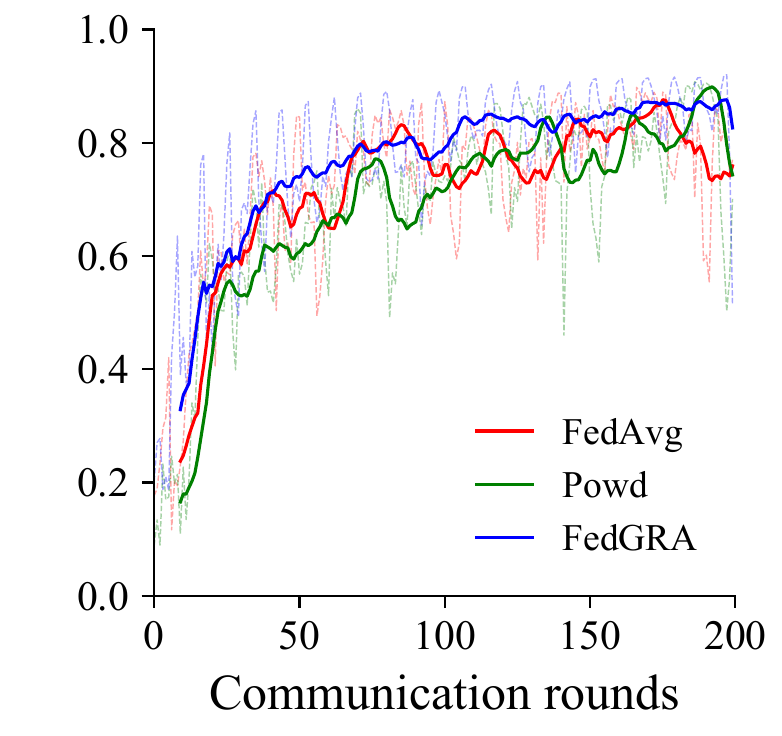}
        \caption{Learning rate = 0.125}
        \label{fig:learning_rate_a}
    \end{subfigure}
    \begin{subfigure}[b]{0.24\textwidth}
        \includegraphics[width=45 mm, height=42.5 mm]{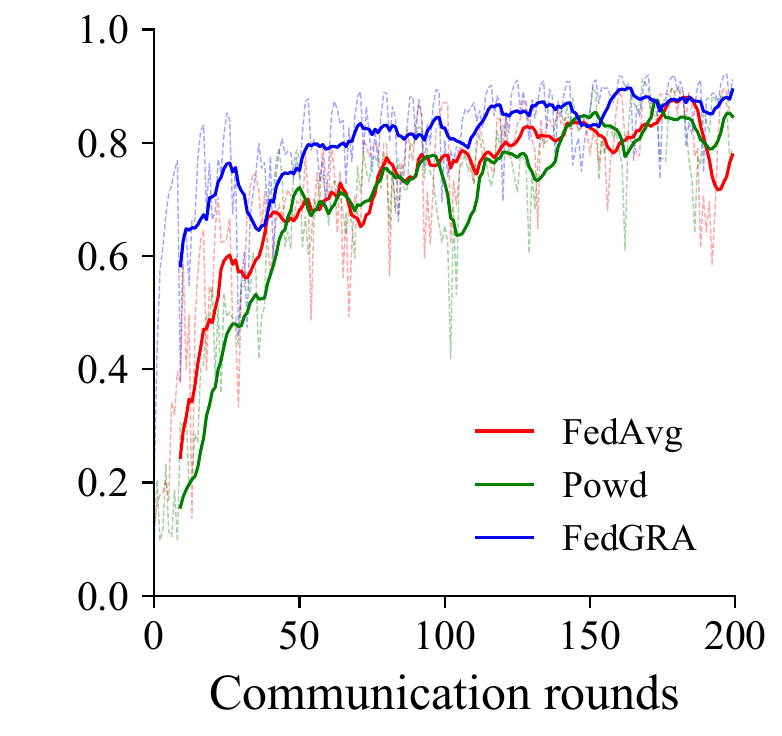}
        \caption{Learning rate = 0.15}
        \label{fig:learning_rate_b}
    \end{subfigure}
    \caption{Evaluating the test set accuracy of FedGRA for different learning rates by using \textbf{MNIST} dataset.}
    \label{fig:learning_rate}
\end{figure}

Fig. \ref{fig:learning_rate} represents the results of the experiments evaluate the impact of changing the learning rate. In the experiment, FedGRA demonstrates consistent and stable convergence compared to FedAvg. When the learning rate is set to 0.125, the accuracy of FedGRA reaches 80\% and 90\% at the 27th and the 117th aggregation rounds, respectively. In contrast, FedAvg takes 50 rounds to reach 80\% accuracy and converges to 90\% accuracy at the 167th round. 


\subsubsection{Number of selected clients} 
\begin{figure}[t]
    \centering
    \begin{subfigure}[b]{0.24\textwidth}
        \includegraphics[width=45 mm, height=42.5 mm]{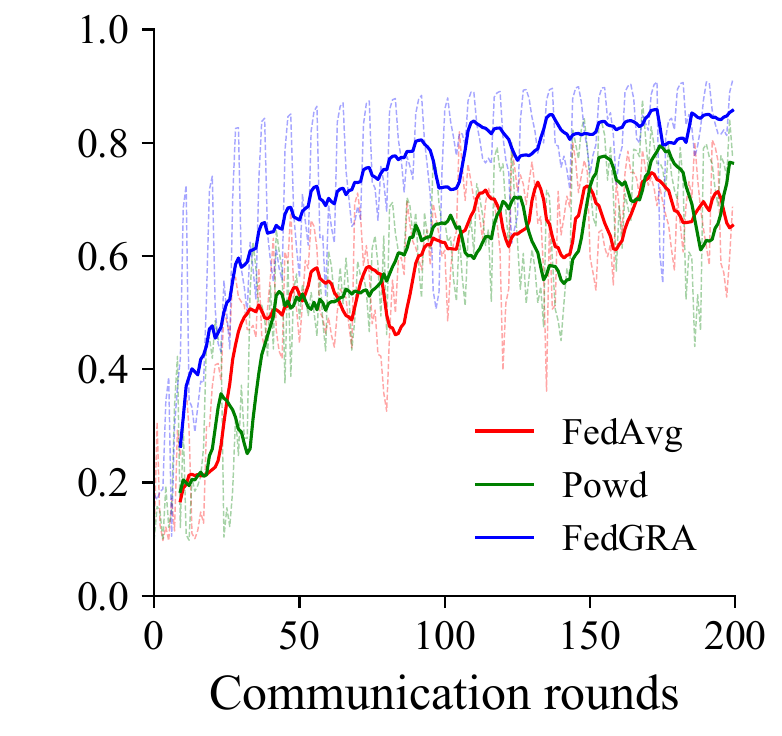}
        \caption{Selected clients = 5}
        \label{fig:selected_client_a}
    \end{subfigure}
    \begin{subfigure}[b]{0.24\textwidth}
        \includegraphics[width=45 mm, height=42.5 mm]{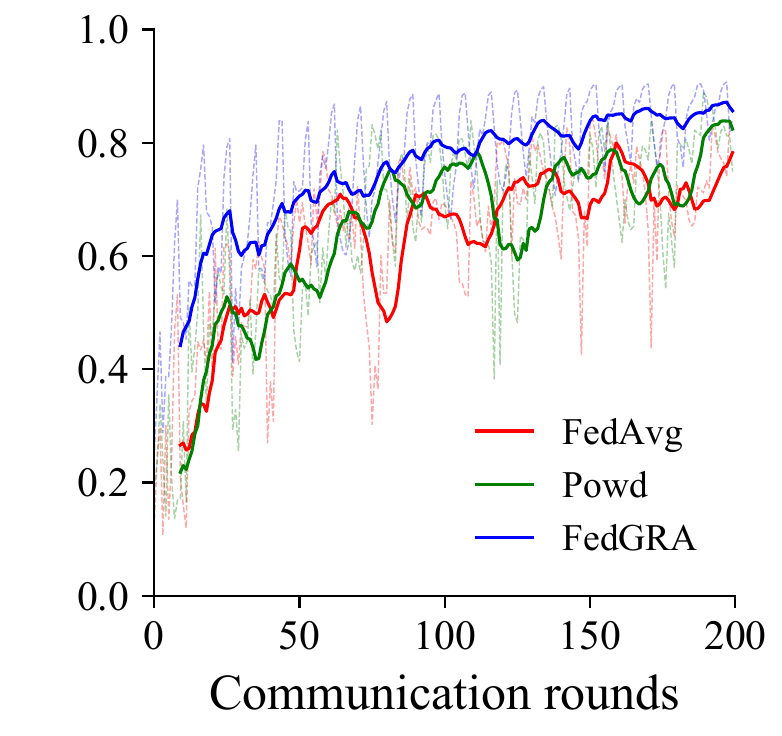}
        \caption{Selected clients = 7}
        \label{fig:selected_client_b}
    \end{subfigure}
    \caption{Evaluating the test set accuracy of FedGRA for varied number of selected clients by using \textbf{MNIST} dataset.}
    \label{fig:selected_client}
\end{figure}

In FL, the accuracy of the global model is heavily affected by the number of clients selected to participate in training for each round. Fig. \ref{fig:selected_client} evaluates the accuracy of FedGRA for different numbers of selected clients. As shown in this figure, the result of FedGRA is much more stable than FedAvg in all cases as it shows fewer accuracy drops. The performance gap between the two methods is increased by reducing the number of clients.

\subsubsection{Client selection round} 
\begin{figure}[h!]
    \centering
    \begin{subfigure}[b]{0.24\textwidth}
        \includegraphics[width=45 mm, height=42.5 mm]{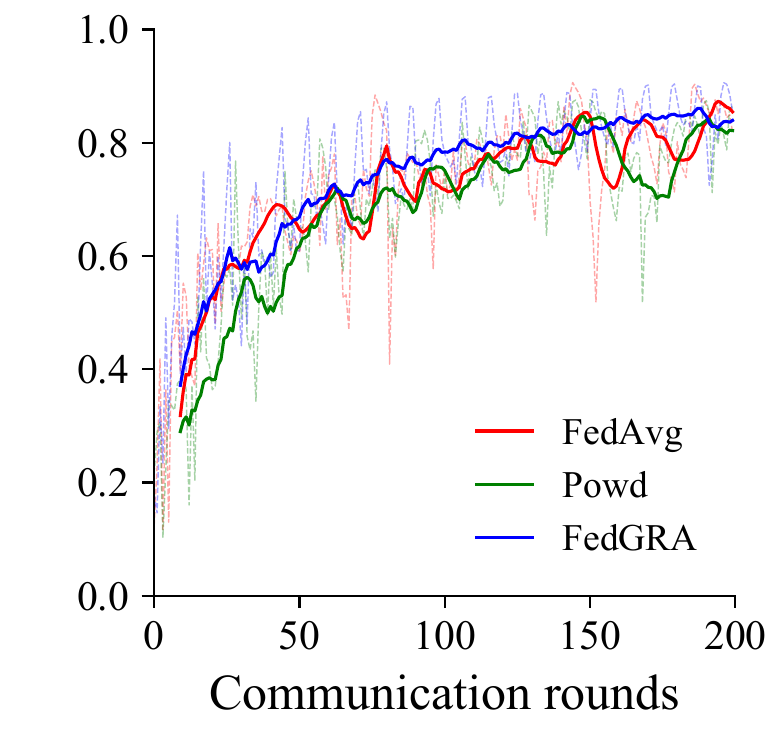}
        \caption{Selection round = 3}
        \label{fig:selection_round_a}
    \end{subfigure}
    \begin{subfigure}[b]{0.24\textwidth}
        \includegraphics[width=45 mm, height=42.5 mm]{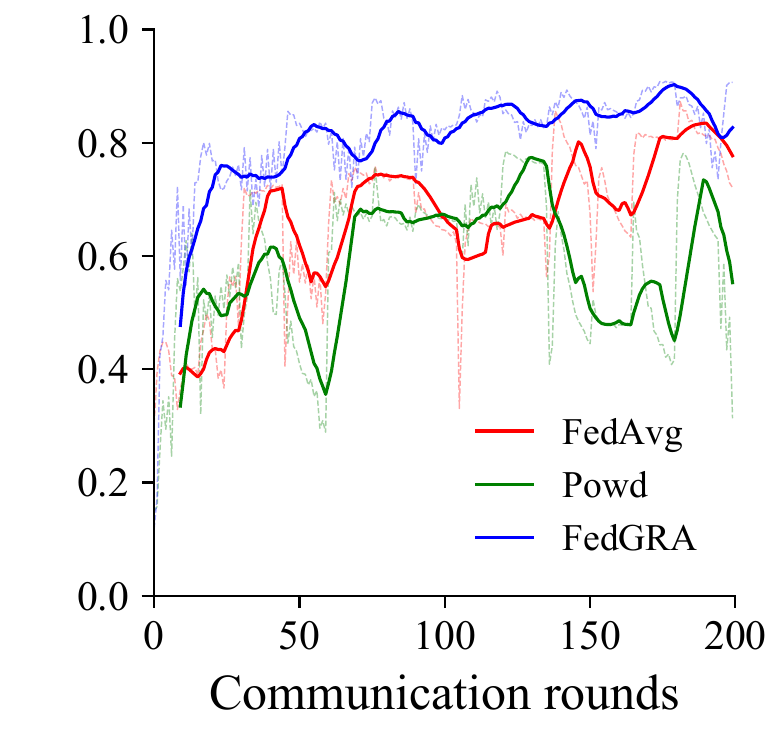}
        \caption{Selection round = 15}
        \label{fig:selection_round_b}
    \end{subfigure}
    \caption{Evaluating the test set accuracy of FedGRA for different client selection rounds by using \textbf{MNIST} dataset.}
    \label{fig:selection_round}
\end{figure}

Reducing the frequency of performing client selection decreases the participation rate of most clients, resulting in a diminished generalization across all classes in the global model, particularly for non-IID data. Fig. \ref{fig:selection_round} evaluates the performance under different client selection rounds. Based on the results, \textbf{FedGRA outperforms FedAvg significantly in terms of overall convergence, with a notable reduction in accuracy fluctuation}. This improvement is attributed to the loss and weight divergence metrics used by FedGRA to enhance the model's generalization across all classes. Additionally, by considering fairness, FedGRA ensures that classes used by less capable clients are trained under a minimum client participation rate.

\section{Conclusion}\label{sectionconclusion}
In this paper, we presented FedGRA, a grey relation analysis theory-based client selection method designed to handle data and device heterogeneity for FL. FedGRA leverages weight divergence and training loss to identify clients that can speed up the convergence of the global model during training under a high level of data non-IIDness. To assess FedGRA's effectiveness, we implemented our proposed method by using TensorFlow and evaluated its performance on a testbed of 50 AWS EC2 instances configured with the most common hardware configurations of mobile devices. By using MNIST and FMNIST datasets, we evaluated the performance of our contribution through a comprehensive series of experiments. Our evaluation results for MNIST and FMNIST datasets indicate that FedGRA significantly improves training efficiency by enhancing the convergence of the global model and also reducing the average client's waiting time in a round of training, compared to the state-of-the-art methods.

\bibliography{citation.bib}
\bibliographystyle{IEEEtran}
\end{document}